\pdfoutput=1
\documentclass[
]{ceurart}

\usepackage{listings}
\begin{document}

\copyrightyear{2023}
\copyrightclause{Copyright for this paper by its authors.
  Use permitted under Creative Commons License Attribution 4.0
  International (CC BY 4.0).}

\conference{HHAI-WS 2023: Workshops at the Second International Conference on Hybrid Human-Artificial Intelligence (HHAI), June 26—27, 2023, Munich, Germany}

\title{`SOS TUTORÍA UC': A Diversity-Aware Application for Tutor Recommendation Based on Competence and Personality}


\author[1]{Laura Achón}[%
orcid=0000-0003-3632-5985,
email=laura.achon@hotmail.com,
]
 
\address[1]{DEI - Universidad Católica Nuestra Señora de la Asunción - Paraguay}
\address[2]{Aalborg University - Denmark}

\author[1]{Ana De\ Souza}[%
orcid=0000-0003-1914-4496,
email=ana.desouzag@outlook.com,
]

\author[1]{Alethia Hume}[%
orcid=0000-0002-1874-1419,
email=alethia.hume@uc.edu.py,
]

\author[2]{Ronald Chenu-Abente}[%
orcid=0000-0002-1121-0287,
email=ronaldchenu@gmail.com,
]

\author[2]{Amalia De\ Götzen}[%
orcid=0000-0001-7214-5856,
email=ago@create.aau.dk,
]

\author[1]{Luca Cernuzzi}[%
orcid=0000-0001-7803-1067,
email=lcernuzz@uc.edu.py,
]


\begin{abstract}
`SOS TUTORÍA UC' is a student connection application aimed at facilitating academic assistance between students through external tutoring outside of the application. To achieve this, a responsive web application was designed and implemented, integrated with the \textit{WeNet} platform, which provides various services for user management and user recommendation algorithms. 
This study presents the development and validation of the experience in the application by evaluating the importance of incorporating the dimension of personality traits, according to the \textit{Big Five} model, in the process of recommending students for academic tutoring. The goal is to provide support for students to find others with greater knowledge and with a personality that is ``different'', ``similar'' or ``indifferent'' to their own preferences for receiving academic assistance on a specific topic. 
The integration with the \textit{WeNet} platform was successful in terms of components, and the results of the recommendation system testing were positive but have room for improvement. 
\end{abstract}


\maketitle

\section{Introduction}
This paper presents `SOS TUTORÍA UC', an application developed by Universidad Católica Nuestra Señora de la Asunción (UC) in Paraguay and integrated into the \textit{WeNet} platform, which aims to facilitate connections between students for tutoring based on academic competencies and personality, utilizing the \textit{Big Five} model \cite{B5} to measure personality traits. Additionally, the application uses hybrid machine artificial intelligence, leveraging the information of the personality traits and the level of academic competencies of potential tutors to make recommendations. This allows users to choose tutors with ``different'' or ``similar'' personalities to their own, and they can also choose to disregard this parameter if they feel ``indifferent'' towards it. The study explores how these personality traits can influence tutoring effectiveness, in addition to academic knowledge in the specific subject to be taught. This is in line with various previous studies that have examined the impact of peer-to-peer tutor recommendation systems \cite{Labarthe, Saad, Potts,Ma2} and personality-based tutor recommendation systems \cite{Niehoff, Menges} on the educational process. Furthermore, the current research builds upon previous studies \cite{Hume, Mercado} that have shown a correlation between personality traits and interaction levels in \textit{WeNet} project pilots, highlighting the importance of diversity in social interactions. 
\section{`SOS TUTORÍA UC'}
The `SOS TUTORÍA UC' web application was designed and implemented, providing an interface for students to request tutoring sessions and find suitable tutors based on their academic expertise and personality traits. 
To incorporate personality into the matching process, we adopted the \textit{Big Five} model \cite{B5}, that uses the traits, namely Extraversion, Agreeableness, Conscientiousness, Emotional Stability/Neuroticism, and Openness to Experience, to provide a comprehensive framework for describing and understanding individual differences in personality.
The `SOS TUTORÍA UC' application was integrated with the \textit{WeNet} platform \cite{Rosell}, ensuring good user management and leveraging the platform's recommendation algorithms.
\begin{figure} [h]
    \centering
    \includegraphics [scale=0.45] {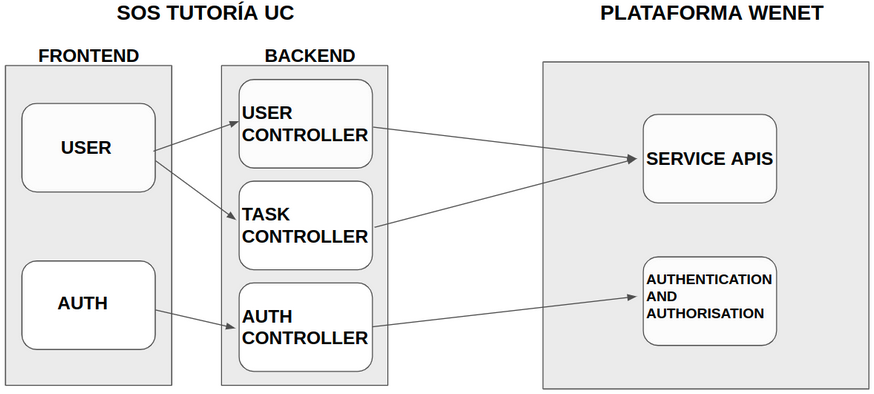}
    \caption{Communication between the modules of 'SOS TUTORÍA UC' and the \textit{WeNet} platform}
    \label{fig:Componentes}
\end{figure}
Figure \ref{fig:Componentes} showcases the integration process, which involved analyzing the components and APIs of the \textit{WeNet} platform. The goal was to create and adapt the logic for tutoring requests to align with the existing \textit{WeNet} model. The Service REST API manages model entities and communicates with modules such as the Profile Manager (user management) and the Task Manager (handling task and message creation as transactions).

In 'SOS TUTORÍA UC,' the creation of a task corresponds to a new tutoring request. The \textit{WeNet} platform provides five recommended users as potential tutors, and they receive a notification to review the request details. Based on the information, they can accept or reject the request for tutorship. The responses to these requests are recorded as task response transactions, which can be approval or rejection responses determined by an attribute value. Once a request is approved by at least one student, the requester can select the desired tutor through a transaction type indicating the best response. This informs the platform which response sent by the tutors has been ultimately chosen by the requesting student.

As part of the integration, the application logic had to be adapted on the rules side as well. The relevant dimensions that contribute to diversity in the algorithm are academic competence, personality, and physical proximity. Finding individuals based on physical proximity is determined by their distance from the person who posted the question. A distance greater than 500 meters is considered far and out of reach, while a shorter distance is classified as closer. Competence is also considered a strict constraint. The system only selects individuals who are ``better than me'' in terms of academic competence, if they are available. Finding individuals with ``similar'' or ``different'' personality profiles is based on whether they share a ``similar'' personality traits with the person requesting tutoring or not. Competence and physical proximity requirements are considered essential and given more importance than other requirements (such as personality). However, it is acceptable to ``diversify'' the list of potential tutors by selecting someone who is not physically close in order to diversify based on gender.

\section{Analysis of results}
To evaluate the system, a pilot study was conducted at the UC. Participants were invited to use the `SOS TUTORÍA UC' application and provide feedback on their experience. Since this work corresponds to the integration of a new type of technology (i.e., a responsive web application) into the \textit{WeNet} platform, differentiating it from other consortium pilots, a component-level integration analysis was conducted. It involved evaluating the communication through the interfaces of the \textit{WeNet} platform; that is, the Service REST APIs in the management of transactions and users, as well as the integration with the authentication and authorization module. The tests were performed with a test user in production environment to the three main modules (Auth Controller, User Controller and Task Controller). All calls to the \textit{WeNet} APIs during the tests returned a satisfactory response, in addition to executing the expected results. Therefore, the component-level integration was successful.

Additionally, we also discuss a preliminary experience with the recommendation process. For this purpose, a testing scheme was established to analyze whether the user profiles suggested by \textit{WeNet} to fulfill a tutoring request adequately match the parameters indicated by the requester. In the tests performed for applications with personality ``different'', ``similar'' or ``indifferent'' to the applicant, the algorithm prioritized students with high scores in the requested competency and the specified personality preference. However, it is worth noting that for some of the tests, the algorithm should have recommend other tutors that better met the applicant's requirements. Indeed, despite the recommendation algorithm yielded positive results, further improvements are needed to better align with user preferences and selection filters specified. This could be attributed to low student participation during the pilot experience or the platform's gender diversification approach.

In the pilot experience participated 43 students from different campuses and careers at the UC. After completing the pilot of `SOS TUTORÍA UC' three primary instruments were used to analyze the experience and evaluate the relevance of incorporating personality as a search parameter for potential tutors, thereby introducing an additional element of diversity: i) questionnaires for tutors and requesters after their connection; ii) an exit questionnaire for the pilot; and iii) a focus group.

The pilot study revealed valuable insights from the participants. Mainly of the suggestions arose from the focus group in which participated 7 students in a hybrid meeting (3 in person and 4 virtually). They expressed the importance of being able to choose tutors based on personality traits, as this can significantly impact the dynamics and effectiveness of the tutoring relationship. Participants suggested the inclusion of additional information, such as the ``level of compatibility'' or ``level of personality traits'', to aid in the selection process and provide a more informed decision-making framework.

One notable challenge observed during the pilot study was the low usage of the `SOS TUTORÍA UC' application. This was attributed to the limited availability of participants during the exam period, which resulted in a reduced number of tutors available to provide support. Consequently, there was a decreased incentive for students to request tutoring sessions through the application. To address this issue, participants recommended extending the usage of the application to the entire academic semester, allowing for more opportunities for tutoring and increasing overall engagement.

Based on the results obtained and the feedback from participants during the focus group, several priority areas for future work were identified. Firstly, there is a need to fine-tune the recommendation algorithms to better align with user preferences and selection filters. This could involve incorporating additional parameters related to personality traits, compatibility levels or other diversification elements. Secondly, developing a mobile application with push notifications for events could enhance the user experience and interaction, making it more convenient for students to engage with the system. Lastly, conducting a longer-term study that covers an entire academic semester would provide more data and insights into the student's experience with the personality filtering feature.

\section{Discussion and conclusions}
This paper presented the incorporation of personality traits based on the \textit{Big Five} model into a peer tutoring system, `SOS TUTORÍA UC', aimed at providing academic support. The integration with the \textit{WeNet} platform successfully enabled the matching of students with tutors based on ``compatible'', ``complementary'' or ``indifferent'' personalities. The pilot study yielded positive results in terms of system integration, and the recommendation algorithm showed favorable outcomes by prioritizing the academic level in making recommendations. However, there is still potential for improvement concerning the consideration of potential tutor's personality traits.

The feedback from participants highlighted the importance of considering personality traits in the tutor-tutee matching process. They expressed the desire for additional information and parameters to facilitate the selection of tutors, such as compatibility levels and more detailed personality trait profiles. The limited usage during the exam period underscored the need to extend the application's availability throughout the academic semester.
\begin{acknowledgments}
This research received funding from the proactive FET Horizon 2020 project of the European Union, \textit{WeNet: Internet of Us}, Grant Agreement No. 823783. We would also like to express our gratitude to the various institutions associated with the \textit{WeNet: Internet of Us} project for their valuable cooperation.
\end{acknowledgments}


\end{document}